\documentclass[prd,twocolumn,nofootinbib,notitlepage]{revtex4-1}
\usepackage{enumitem}
\usepackage{graphicx}
\usepackage{subfigure}
\usepackage{amssymb}
\usepackage{multirow}
\usepackage{amsmath}
\usepackage{bbm}
\usepackage{color}
\usepackage[retainorgcmds]{IEEEtrantools}

\newcommand{\beq}{\begin{equation}}
\newcommand{\eeq}{\end{equation}}
\newcommand{\ignore}[1]{}
\newcommand{\be}{\begin{equation}} \newcommand{\ee}{\end{equation}}
\newcommand{\bea}{\begin{eqnarray}} \newcommand{\eea}{\end{eqnarray}}

\begin{document}
\title{New Limit on Pseudoscalar-Photon Mixing from WMAP Observations}
\author{Prabhakar Tiwari}
\date{30 Oct 2012}
\email{ptiwari@iitk.ac.in}
\affiliation{
Department of Physics, Indian Institute of Technology, Kanpur - 208016, India\\
}
\begin{abstract}
The pseudoscalar-photon mixing in presence of large scale magnetic 
field induces polarization in light from distant cosmological sources. We 
study the effect of these pseudoscalars or axion like particles (ALPs) 
on Cosmic Microwave Background Radiation (CMBR) and constrain the product 
of mixing strength $g_{\phi}$ times background magnetic field $B$. The 
background magnetic field has been assumed to be primordial 
and we assume large scale correlations with the correlation length of 1 Mpc. 
We use WMAP seven year foreground reduced polarization and temperature data 
to constrain pseudoscalar-photon mixing parameter. We look for different mass 
limits of the pseudoscalars and find $g_{\phi}B\le 1.6\times10^{-13} GeV^{-1} nG$ 
with ALPs of mass $10^{-10} eV$ and $g_{\phi}B\le3.4\times10^{-15} GeV^{-1} nG$
for ultra light ALPs of mass $10^{-15} eV$. 
\end{abstract}

\maketitle

\section{Introduction}
\label{one}
The pseudoscalar-photon mixing and its effects on distant cosmological sources
 have been studied in literature
\cite{Harari:1992,Mohanty:1993,Das:2001,Kar:2002,Kar:2002cqg,Csaki:2002,Csaki:2002prl,Grossman:2002,Jain:2002vx,
Song:2006,Mirizzi:2005,Raffelt:2008,Gnedin:2007,Mirizzi:2007,Finelli:2009,Agarwal:2012,Ostman:2005,Lai:2006,
Hooper:2007,Hochmuth:2007,Chelouche:2009}.
These hypothetical axion like particles (ALPs), arise naturally as pseudo-Goldstone bosons in theories with 
spontaneously broken global symmetries
\cite{Peccei:1977prl,Peccei:1977prd,McKay:1977,Weinberg:1978,Wilczek:1978,
McKay:1979,Kim:1979,Dine:1981,Kim:1987}. ALPs have an interaction vertex with  two photons
and hence in an external magnetic field ALPs can convert into a photon and vice versa
\cite{Clarke:1982,Sikivie:1983,Sikivie:1985,Sikivie:1988,Maiani:1986,Raffelt:1988,Carlson:1994,Bradley:2003,
Das:2004qka,Das:2004ee,Ganguly:2006,Ganguly:2009}.
Although this effect is very small,
it becomes significant at cosmological scales and leads to many interesting signatures on electromagnetic 
radiation. This pseudoscalar-photon mixing phenomena causes  changes in intensity as well as 
polarization in radiation from  distant sources
\cite{Harari:1992,Mohanty:1993,Das:2001,Kar:2002,Kar:2002cqg,Csaki:2002,Csaki:2002prl,Grossman:2002,Jain:2002vx,
Song:2006,Mirizzi:2005,Raffelt:2008,Gnedin:2007,Mirizzi:2007,Finelli:2009,Agarwal:2012,Ostman:2005,Lai:2006,
Hooper:2007,Hochmuth:2007,Chelouche:2009}. The contribution of this effect has been investigated for CMBR
\cite{Mirizzi:2005,Agarwal:2008ac}, radio \cite{Harari:1992,Jain:2002vx,Jain:1998r,Ralston:2004} and optical 
\cite{Agarwal:2012,Agarwal:2011,Hutsemekers:1998,Hutsemekers:2001fv,Hutsemekers:2005iz,Hutsemekers:2008} sources.
Various experiments are looking for these pseudoscalars and providing limit on the coupling constant $g_{\phi}$ and 
their masses $m_{\phi}$\cite{Mohanty:1993,Raffelt:2008,Dicus:1978,Dearborn:1986,Raffelt:1987,Raffelt:1988prl,Turner:1988,
Janka:1996,Keil:1997,Brockway:1996,Grifols:1996,Raffelt:1999,Rosenberg:2000,Horns:2012,Zioutas:2005,
Lamoreaux:2006,Yao:2006,Jaeckel:2007,Andriamonje:2007,Robilliard:2007,Zavattini:2008,Rubbia:2008}.
 
In the present paper we study the effect of pseudoscalar-photon mixing on CMBR multipoles.
We show, using WMAP observations that this leads to a new constraint on the product of magnetic field 
$B$ and the pseudoscalar-photon coupling $g_{\phi}$. 
We consider the background as a large number of correlated magnetic field domains 
and do a complete $3D$-simulation to calculate the 
Stokes parameters for CMBR. The origin of background magnetic 
field is considered as primordial \cite{Subramanian:2003sh,Seshadri:2005aa,
Seshadri:2009sy,Jedamzik:1998,Subramanian:1998} and we assume a smooth variation of the 
magnetic field over the scale of 1 Mpc. The magnetic field correlations are assumed 
to obey a power law with spectral index $n_B$. 
The details for the background magnetic field model are discussed in Sec.\ref{sc:PMF}.
As we do simulation over a very large distances (redshift 1000), 
we choose domain size around 16 Mpc. The strength of magnetic field in each domain is assumed to be order of 
nG \cite{Csaki:2002prl,Grossman:2002,Mirizzi:2005}. 

 The initial pseudoscalar density is assumed to be zero or negligible as compared
to photon density as assumed by most authors \cite{Das:2004ee,Agarwal:2008ac,Agarwal:2012,Agarwal:2011}.
We made this assumption as the pseudoscalars are likely to decouple from cosmic plasma 
at very early times. After pseudoscalar decoupling, photon density would be enhanced by 
many processes such as QCD phase transition, $e^- e^+$ annihilation etc. It may not even be in equilibrium 
after inflation. Hence, it is reasonable to assume pseudoscalar density as negligible as compared
to photon density.  

We compare our result with the WMAP 7-year data
and constrain the coupling parameter $g_{\phi}$ times $B$. The limit presented in 
the paper is bound to certain assumptions. We list all of them as follows:\\
1)The background magnetic field follows a simple cosmological evolution.\\
2)We have assumed a definite value for the spectral index $n_{B}=-2.37$ which correspond to 
the best fit of matter and CMBR power spectrum\cite{Yamazaki:2010nf}. However we also determine 
its dependence on $n_{B}$.  \\
3)CMBR is assumed to be unpolarized initially and the initial density for pseudoscalars is zero.\\

The paper is organized as follows. In Sec.\ref{sc:mixing} we briefly review the pseudoscalar-photon mixing in presence of 
plasma and uniform magnetic field in a flat expanding universe. In Sec.\ref{sc:PMF} we model the background 
magnetic field, which is correlated in real space and discuss the numerical method for generating the 
$3D$ magnetic field. In Sec.\ref{sc:sim} we present our simulation result and 
compare with the WMAP observations. Finally, in Sec.\ref{sc:dis} we conclude and compare our results 
with available literatures.   
\section{pseudoscalar-photon mixing}
\label{sc:mixing}
\subsection{Basic Formulation}
In this section we briefly describe the propagation of electromagnetic 
waves coupled to a pseudoscalar field. 
The basic action for the coupling of pseudo-scalar field $\phi$ to electromagnetic field in the flat
expanding universe is given as \cite{Agarwal:2012,Carroll:1991,Garretson:1992},
\begin{IEEEeqnarray}{rCl}
S= \int d^{4}x \sqrt{-g}~ \Big[-\frac{1}{4}F_{\mu\nu}F^{\mu\nu}-\frac{1}{4} g_{\phi} \phi F_{\mu\nu}\tilde{F}^{\mu\nu}
\nonumber\\+\frac{1}{2}(\omega_{p}^{2}a^{-3})A_{\mu}A^{\mu}+\frac{1}{2}g^{\mu\nu}\phi_{,\mu}\phi_{,\nu}-
\frac{1}{2}m^{2}_{\phi}\phi^{2}\Big].
\label{eq:S_expanding}
\end{IEEEeqnarray}
Here $F_{\mu\nu}$ is the electromagnetic field tensor and $\tilde{F}^{\mu\nu}$ the dual tensor,
$g_{\phi}$ is the  coupling constant between  $\phi$ to photon field and `$a$' the usual cosmological scale
factor. In above action Eq.$(\ref{eq:S_expanding})$, we have a plasma frequency ($\omega_p$) term  as
$\frac{1}{2}(\omega_{p}^{2}a^{-3})A_{\mu}A^{\mu}$, which acts as an effective mass term for
photon. We may note that this term scales as $a^{-3}$($volume^{-1}$), as $\omega_{p}^{2}$ is
proportional to the plasma number density.

We choose a fixed coordinate system such that the z-axis lies along the 
direction of propagation. We define 
$ {\boldsymbol {\cal A}}= \frac{(a^{2}{\boldsymbol E})}{\omega}$ where ${\boldsymbol E}$
is the usual electric field vector and $\omega$ is the radiation frequency. Only the component 
of ${\boldsymbol {\cal A}}$ parallel to $B$ transverse mixes with the pseudoscalar field $\phi$.
We replace $\phi$ by $\frac{\chi}{a}$ and  
the mixing of ${\cal A}_{\parallel}$ to $\chi$ can be written as,
\beq
\label{eq:mixing}
(\omega^{2}+ \partial^{2}_{z})\left(\begin{array}{c}
{\cal A}_{\parallel}\\ \chi \end{array} \right) -
M \left( \begin{array}{c} {\cal A}_{\parallel}\\ \chi \end{array} \right) =0, 
\eeq
where M is the `mixing matrix' as,
\beq
\label{eq:matrix} 
M = \left(\begin{array}{cc}

         \frac{\omega^{2}_{p}}{a}     & ~~~ -\frac{g_{\phi}}{a^{2}}
(a^{2}\mathcal{B}_\perp)\omega\\
-\frac{g_{\phi}}{a^{2}}\ (a^{2}\mathcal{B}_\perp)\omega   & ~~~ m^{2}_{\phi}a^{2} \end{array}\right).
\eeq

Here $m_{\phi}$ is the pseudoscalar mass, $(a^{2}\mathcal{B}_\perp)$ is the transverse component of magnetic 
field and the factor $a^{2}$ is scaling the magnetic field in expanding universe model.
 
 We follow the procedure described in Ref.\cite{Das:2004ee,Agarwal:2008ac} for solving Eq.\ref{eq:mixing}. 
\subsection{Propagation and Polarization}
In this section we briefly review the propagation of the 
mixed $\cal A$ and $\chi$ field in presence of an external magnetic field.
A detailed description is given in Ref.\cite{Das:2004ee,Agarwal:2008ac}.
We start with some initial densities of $\cal A$ and $\chi$ mixed fields and calculate 
the same after a propagation of distance $z$. An appropriate and general representation 
is given as, 
\beq
\rho(z)= P(z)\rho(0)P(z)^{-1},
\eeq
where $\rho(0)$ is the general density matrix for the fields and P(z) is a non trivial 
unitary matrix, describing the solution for the mixed fields. The representation for $\rho(0)$ and $P(z)$ for 
our coordinate system is as follows,  
\begin{widetext}
\beq 
\rho(0) = \left(\begin{array}{ccc}
<{\cal A_{\parallel}}(0){\cal A_{\parallel}^{\ast}}(0)>    & <{\cal A_{\parallel}}(0){\cal A_{\perp}^{\ast}}(0)>   & <{\cal A_{\parallel}}(0) \chi^{\ast}(0)> \\
<{\cal A_{\perp}}(0){\cal A_{\parallel}^{\ast}}(0)>    & <{\cal A_{\perp}}(0){\cal A_{\perp}^{\ast}}(0)>   & <{\cal A_{\perp}}(0) \chi^{\ast}(0)> \\
<\chi(0) {\cal A_{\parallel}^{\ast}}(0)>    & <\chi(0){\cal A_{\perp}^{\ast}}(0)>   & <\chi(0) \chi^{\ast}(0)>\end{array}\right).
\eeq
and 
\beq 
P(z) = e^{i(\omega+\triangle_{A})z}\left(\begin{array}{ccc}
1-\gamma sin^2 \theta            &   0                  &         \gamma cos\theta sin\theta \\
0                                &  e^{-i[\omega+\triangle_A -(\omega^2- \omega^2_p)^{1/2}]z}  &          0 \\
\gamma cos\theta sin\theta        &     0                       &      1- \gamma cos^2\theta \end{array}\right).
\eeq
\end{widetext}
where $\gamma=1-e^{i\triangle z}$ , while  $\triangle =\triangle_\phi -\triangle_A$ and $\triangle_\phi$ , $\triangle_A$ 
are defined in terms of $\omega$ and  eigenvalues, $\mu^2_\pm$ of the matrix M,
\beq \triangle_A = \sqrt {\omega^2-\mu^2_{+}}-\omega, ~~~~~~
\triangle_\phi = \sqrt{\omega^2-\mu^2_{-}}-\omega. \eeq
We assume that the CMBR is unpolarized at $z=1000$ and 
set initial density for pseudoscalars 
to be zero. Initially the plasma density is supposed to be very low ($n_{e}=3.24\times10^{-10} a^{-3}cm^{-3}$) \cite{dodelson:MC}
($z \approx 1000$) as the universe has gone through the 
recombination ($z \approx 1100$) era and almost all the electrons 
and protons have been combined to form  neutral hydrogen.
At redshift $z \approx6$\cite{Becker:2001}, the star formation 
starts and the universe becomes ionized, so for  redshift $z<6$ we 
assume  plasma density  $n_{e}=10^{-8} a^{-3}cm^{-3}$. We next propagate the CMBR  through 
these domains and  compute $\rho(z)$ in each domain. Finally we
obtain  $\rho(z)$ and calculate the Stoke's parameters as, 
\bea
I(z)= <{\cal A_{\parallel}}(z){\cal A_{\parallel}^{\ast}}(z)> + <{\cal A_{\perp}}(z){\cal A_{\perp}^{\ast}}(z) >~~~~~~~~\\
Q(z)= <{\cal A_{\parallel}}(z){\cal A_{\parallel}^{\ast}}(z)> - <{\cal A_{\perp}}(z){\cal A_{\perp}^{\ast}}(z) >~~~~~~~~\\
U(z)= <{\cal A_{\parallel}}(z){\cal A_{\perp}^{\ast}}(z)>  + <{\cal A_{\perp}}(z){\cal A_{\parallel}^{\ast}}(z)>~~~~~~~~\\
V(z)= \mathrm{i}(-<{\cal A_{\parallel}}(z){\cal A_{\perp}^{\ast}}(z)>  + <{\cal A_{\perp}}(z){\cal A_{\parallel}^{\ast}}(z)>)~~~
\eea
We compute  multipole anisotropy in E and B modes and constrain $g_{\phi}B$
 by demanding consistency with CMBR observations. 
\section{Background magnetic Field}
\label{sc:PMF}
It is reasonable to assume the origin of background magnetic field as primordial\cite{Mack:2002,Subramanian:2003sh,Seshadri:2005aa,Seshadri:2009sy,Jedamzik:1998,Subramanian:1998}. A two-point correlation 
function for a homogeneous and isotropic magnetic field is given as, 
\bea
        \langle b_{i}({\boldsymbol k}) b^{*}_{j} ({\boldsymbol q}) \rangle & = & \delta_{{\boldsymbol{k,q}}} P_{ij}({\boldsymbol k}) M(k) 
\label{eq:corr}
\eea
where $b_j({\boldsymbol k})$ is the $j^{\rm th}$ component of the
magnetic field in wave vector space. The real space magnetic field
$B_{j}({\boldsymbol r})$ can be written as a Fourier transform of $b_j({\boldsymbol k})${\footnote{%
The real and wave vector space field transformation are as follows,
\bea
\label{eq:IFT}
B_{j}({\boldsymbol r}) &=& \frac{1}{V(2\pi)^{3}} \sum b_{j}({\boldsymbol k}) e^{i {\boldsymbol k}.{\boldsymbol r}},\\
b_{j}({\boldsymbol k}) &=&\frac{1}{V} \sum B_{j}({\boldsymbol r}) e^{-i {\boldsymbol k}.{\boldsymbol r}}.
\eea
}}. Here $P_{ij}({\boldsymbol k}) = \left(\delta_{ij} -\frac{k_{i} k_{j}}{k^2}\right)$ 
is the projection operator and function $M(k)$ is given as, 
\beq M(k)=A k^{n_{B}},\eeq
where $n_{B}$  is power spectral index and the constant A is a normalization.
 The numeric value for A is such as 
$\sum_i <B_i({\boldsymbol r}) B_i({\boldsymbol r})> = B_0^2$ , where $B_0$ is the 
strength of the magnetic field, often assumed to be $1nG$ \cite{Yamazaki:2010nf,Seshadri:2009sy} 
on a comoving scale of 1 Mpc. 
We do not impose any cutoff on correlations in real space and simply choose $r_{max}$ larger than our system.
In other words the lower limit on wave vectors $k_{min}=r_{max}^{-1}$ tends to zero. 

We split the space in $1024\times1024\times1024$ equal
volume domains and generate the $3D$ k-space magnetic field in each domain using the
spectral distribution as in Eq.\ref{eq:corr}.
We use polar coordinate $(k,\theta,
\phi)$ in wave vector space. In k-space the domains are uncorrelated
and for any wave vector ${\boldsymbol k}$, $b_k=0$ and $b_\theta$ and $b_\phi$ are uncorrelated.
Hence, we can generate the $b_\theta$ and $b_\phi$ independently for each domain using a smooth 
Gaussian distribution\cite{Agarwal:2012,Agarwal:2011}. 
\begin{eqnarray}
\label{eq:gauss}
        f(b_{\theta}({\boldsymbol k}),b_{\phi}({\boldsymbol k})) = N \ {\rm exp} 
\left[-\left(\frac{b_{\theta}^{2}({\boldsymbol k}) + b_{\phi}^{2}({\boldsymbol k})}{2M({\boldsymbol k})}\right) \right],
\end{eqnarray}
Here $N$ is a normalization factor. We use this to generate full $3D$ k-space
magnetic field for each domain and do a Fourier transformation to
get the three Cartesian components of the magnetic field in real space.
\section{Simulation and Result}
\label{sc:sim}
We propagate CMBR  from redshift $1000$ and do the simulation for 
the  pseudoscalar-photon mixing in the correlated magnetic field 
background. We perform our computation on a $3D$ grid of $1024\times1024\times1024$. Since the total 
linear distance for redshift $1000$ correspond to a       
very large comoving distance 8104 Mpc (matter dominated universe),
we set one domain size to be 16 Mpc. However, we still keep the correlation 
length fixed to 1 Mpc. This domain size is very large as compared to the oscillation length (0.4 Kpc) 
for the CMBR. The domain size dependence has been studied in Ref.\cite{Agarwal:2012} and a very small statistical
variations have been reported. Hence we do not expect a very significant dependence on the domain size.
We use HEALPix\footnote{http://healpix.jpl.nasa.gov/} to generate angular positions of
the sources with resolution
parameter Nside $=$ 256. 
Next, we  propagate all these sources through each domain and 
determine the Stoke's parameters.

We compare our simulations to seven year foreground reduced WMAP CMBR observations.
We use W-band since it contains the least foreground contamination. 
We demand that the pseudoscalar coupling introduces the temperature fluctuation less that $10^{-5}$.
Furthermore  we demand that the mixing mechanism generates
E and B mode less than or equal to the observed values. In 
our analysis we always have the coupling constant $g_{\phi}$ multiplied with the background magnetic field
$B$ and so we are able to put a limit only on the term $g_{\phi}B$.
We simulate the CMBR polarization with the following parameters:- \\
(1)$B_{T}=1~nG$\\
(2)plasma density $n_{e}=10^{-8} a^{-3}cm^{-3}$ for $z<6$ and \\
$n_{e}=3.64\times10^{-10} a^{-3} cm^{-3}$ for $z\ge6$ \cite{dodelson:MC}\\
(3)CMBR frequency for W-band $=90~GHz$\\
(4) pseudoscalar mass =$ 10^{-10} eV$ and $ 10^{-15} eV$ \\

\begin{figure}[!t]
  \centering{
    \includegraphics[width=3.5in,angle=0]{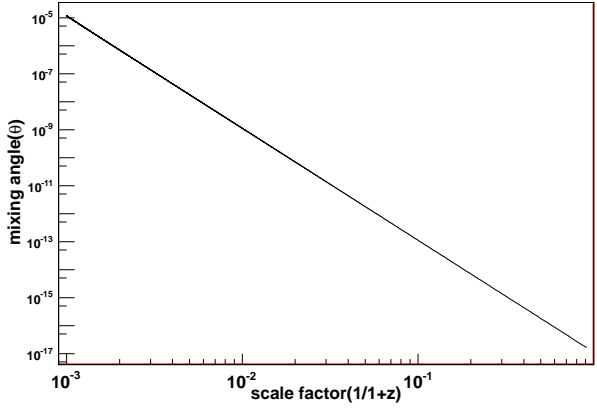}
    \caption{The mixing angle ($\theta$) in flat expanding universe, with parameters 
 $m_{\phi}=10^{-10}eV$, $g_{\phi}B =1.6\times10^{-13}GeV^{-1} nG$.}
\label{fig:ratio1} }
\end{figure}

\begin{figure}[!t]
  \centering{
    \includegraphics[width=3.5in,angle=0]{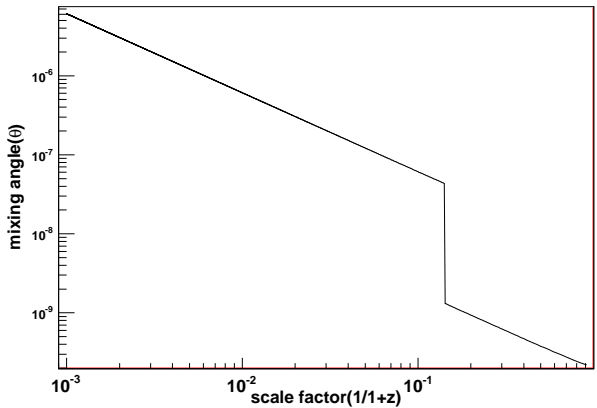}
    \caption{The mixing angle ($\theta$) in flat expanding universe, with parameters  
$m_{\phi}=10^{-15}eV$, $g_{\phi}B = 3.4\times10^{-15} GeV^{-1} nG$. Here plasma mass term (${\omega_p^{2}}/{a}$) is 
dominating over pseudoscalar mass term ($m_{\phi}^{2}a^{2}$).}
\label{fig:ratio2}}
\end{figure}

It turns out that most of the polarization and anisotropy is generated at high redshift. At low
redshift the mixing is negligible as
shown in Fig.(\ref{fig:ratio1},\ref{fig:ratio2}). 
We simulate the mixing effect for $m_\phi = 10^{-10}eV$ and $m_\phi =10^{-15}eV$. 
We find that the pseudoscalar mass is a very significant parameter 
and the predicted CMBR polarization increases rapidly with decreasing pseudoscalar mass unless
it touches the plasma mass limit. For the ultra light pseudoscalar ($m_\phi =10^{-15}eV$) the plasma 
mass term dominate and hence the mixing is controlled by plasma density. This can be seen in Fig.\ref{fig:ratio2}
where we have a sudden dip at $a\sim0.14$, which correspond to reionization era at which the plasma number 
density increases roughly by a factor of $30\sim40$.

\begin{figure}[!t]
  \centering{
    \includegraphics[width=3.5in,angle=0]{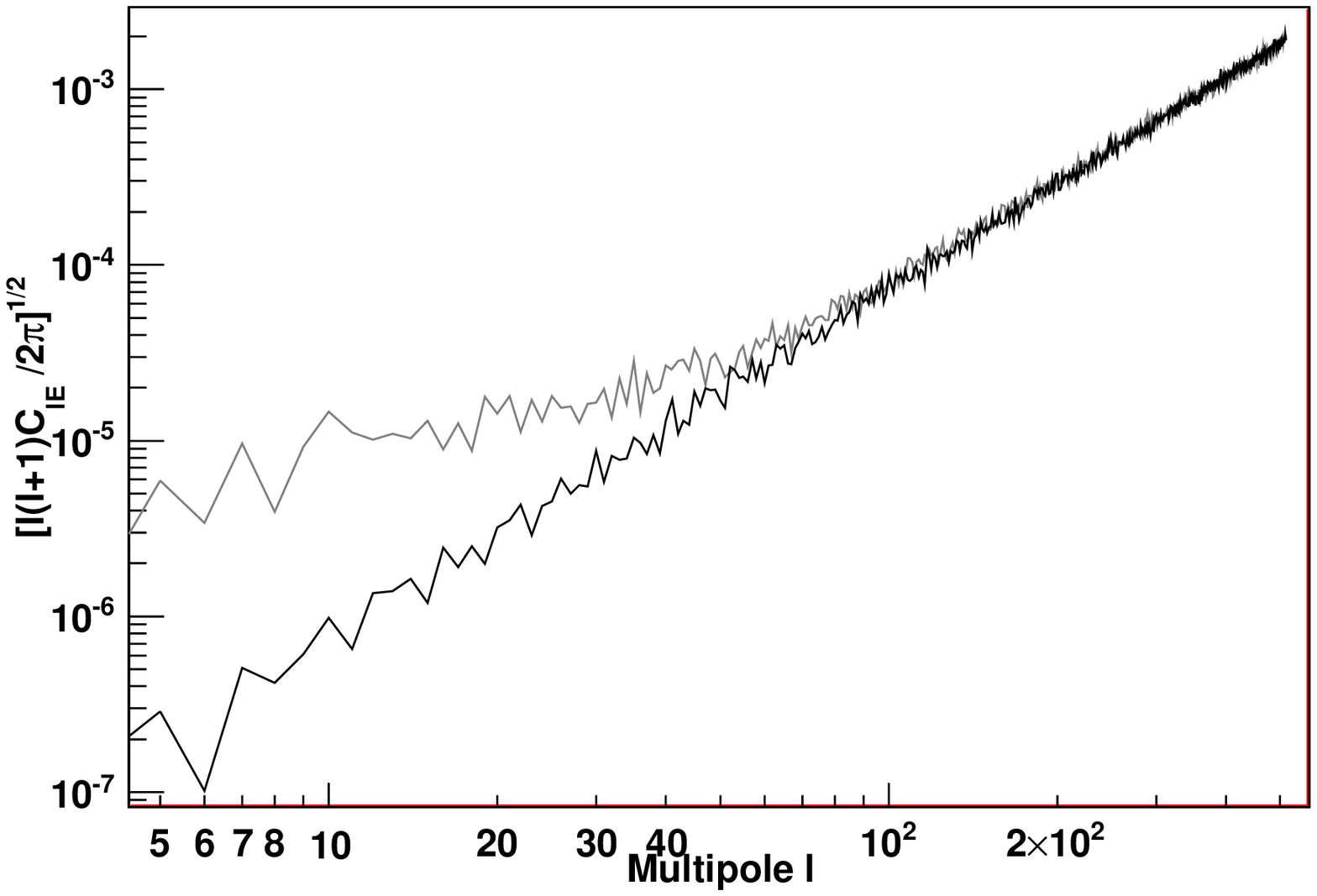}
    \caption{The simulated (black) and WMAP observed (gray) E mode  multipole, using 
$m_{\phi}=10^{-10}eV$, $g_{\phi}B =1.6\times10^{-13}GeV^{-1} nG$. }
\label{fig:Emod1}}
\end{figure}

\begin{figure}[!t]
  \centering{
    \includegraphics[width=3.5in,angle=0]{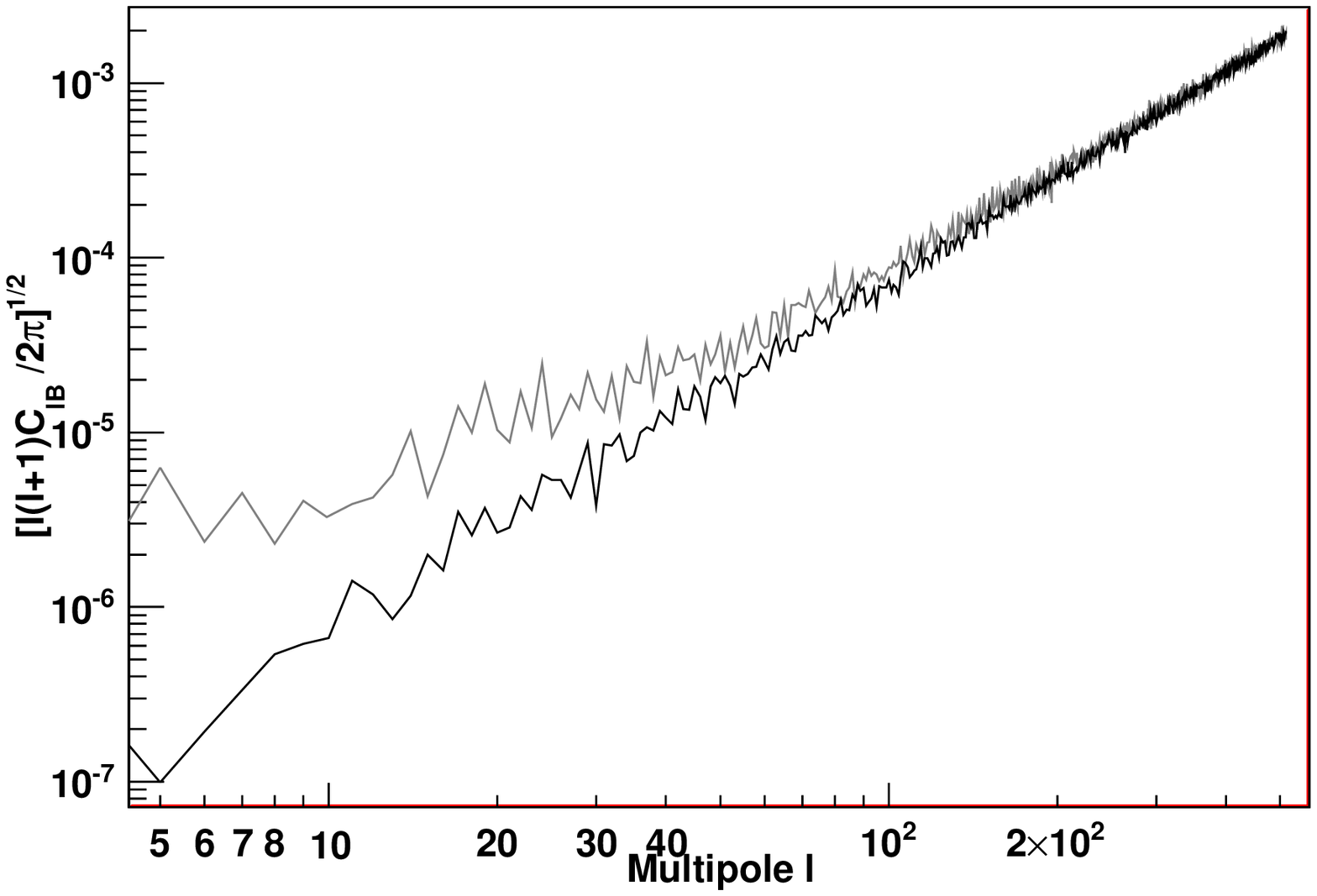}
    \caption{The simulated (black) and WMAP observed (gray) B mode  multipole, using 
$m_{\phi}=10^{-10}eV$, $g_{\phi}B =1.6\times10^{-13}GeV^{-1} nG$.}
\label{fig:Bmod1}}
\end{figure}

\begin{figure}[!t]
  \centering{
    \includegraphics[width=3.5in,angle=0]{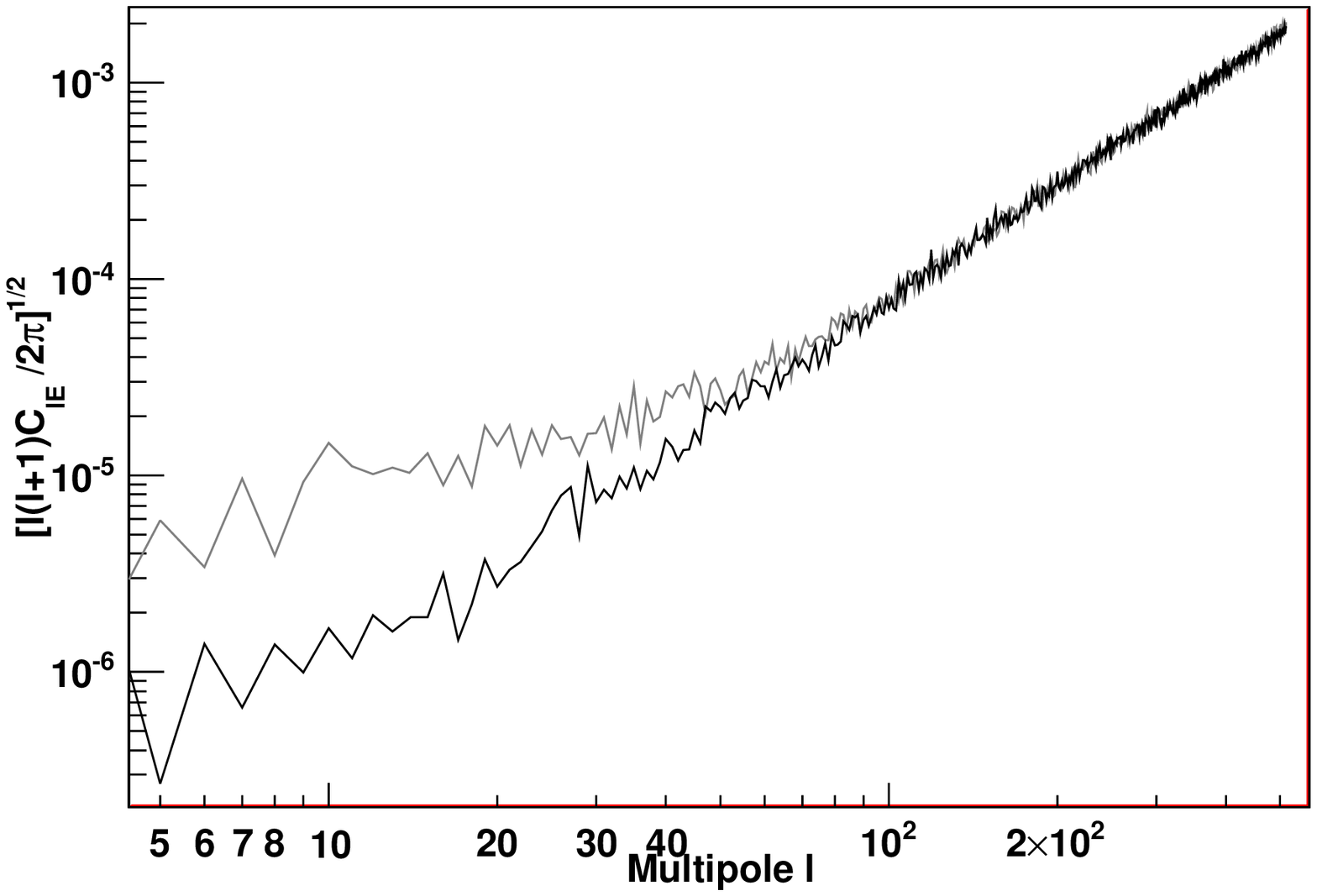}
    \caption{The simulated (black) and WMAP observed (gray)  E mode  multipole, using 
$m_{\phi}=10^{-15}eV$, $g_{\phi}B =3.4\times10^{-15}GeV^{-1} nG$. }
\label{fig:Emod2}}
\end{figure}

\begin{figure}[!t]
  \centering{
    \includegraphics[width=3.5in,angle=0]{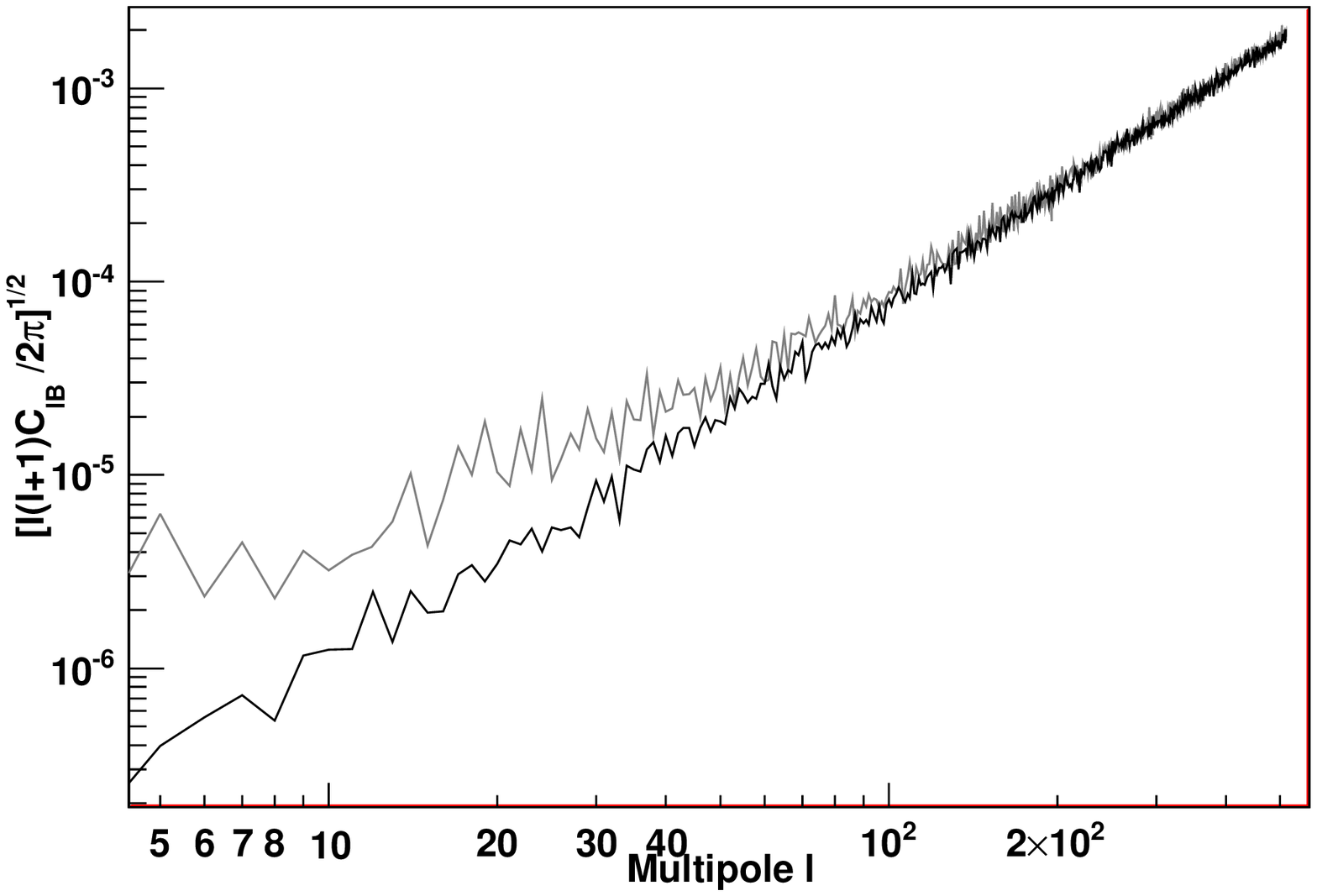}
    \caption{The simulated (black) and WMAP observed (gray) B mode  multipole, using 
$m_{\phi}=10^{-15}eV$, $g_{\phi}B =3.4\times10^{-15}GeV^{-1} nG$.}
\label{fig:Bmod2}}
\end{figure}

\begin{figure}[!t]
  \centering{
    \includegraphics[width=3.5in,angle=0]{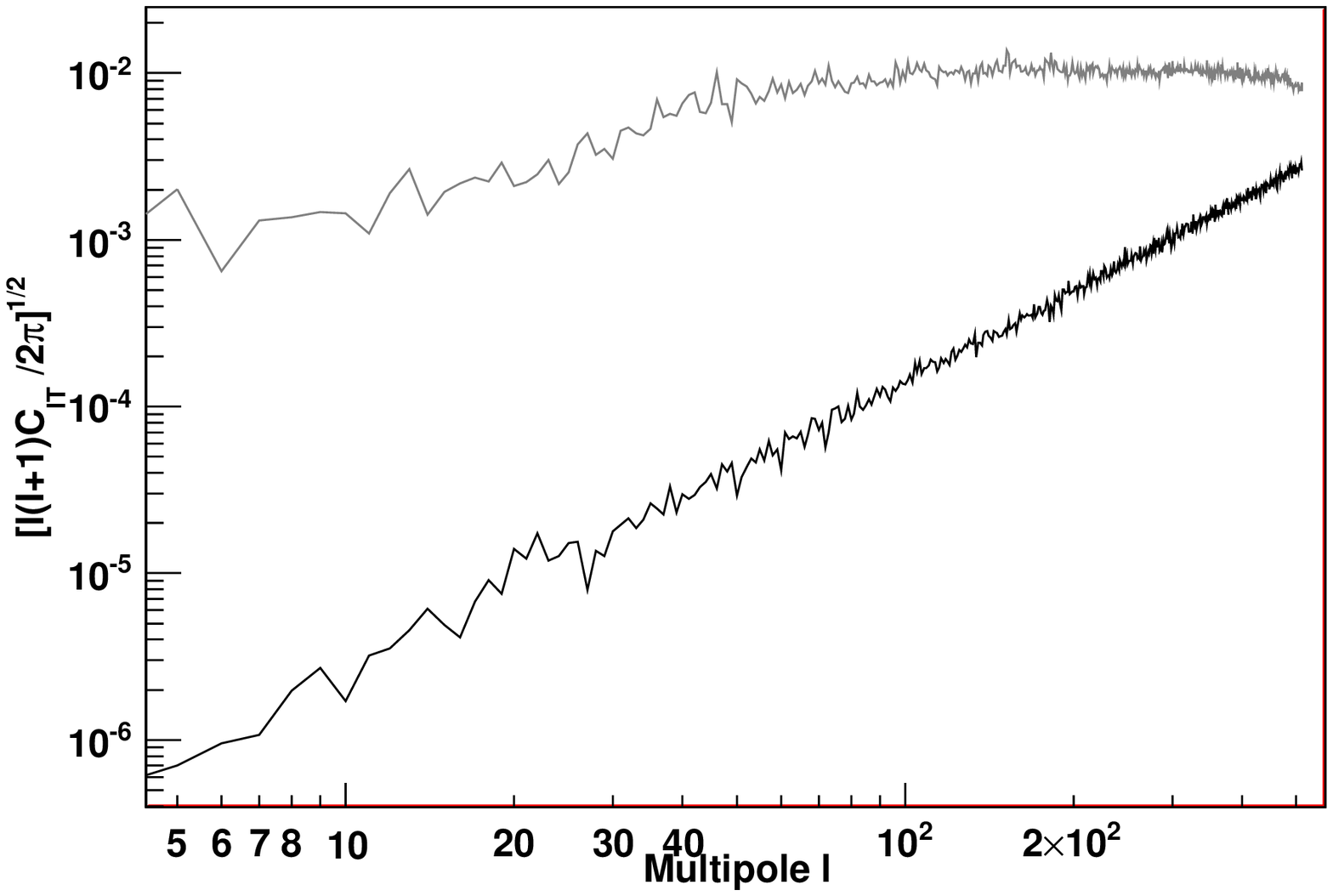}
    \caption{The simulated (black) and WMAP observed (gray)  T mode  multipole,
$m_{\phi}=10^{-10}eV$, $g_{\phi}B =1.6\times10^{-13}GeV^{-1} nG$. }
\label{fig:Tmod1}}
\end{figure}

CMBR fluctuations are analysed by decomposition in terms of spherical harmonics, which allows
us to compute the power in different multipoles. We generate a full sky map of Stoke's 
parameters with 13.7 arcmin resolution for W-band frequency and calculate E and B modes.
We degrade the WMAP seven year foreground reduced I,Q,U sky maps to match with our simulation resolution 
and deduce the E and B modes for W-Band frequency. 
We choose an appropriate value for the factor $g_{\phi}B$ requiring the E and B mode 
multipoles are within the observed value. This leads to a limit on the factor $g_{\phi}B$. We 
show our results in Fig.(\ref{fig:Emod1},\ref{fig:Bmod1}) for $m_{\phi}=10^{-10}eV$
and in Fig.(\ref{fig:Emod2},\ref{fig:Bmod2}) for $m_{\phi}=10^{-15}eV$.
If we set $m_{\phi}=10^{-10}eV$ the limit is fixed to be  $g_{\phi}B\le1.6\times10^{-13}GeV^{-1} nG$.
Alternatively if we choose ultra light pseudoscalar of mass $m_{\phi}\le10^{-15}eV$, 
we obtain $g_{\phi}B\le3.4\times10^{-15}GeV^{-1} nG$. 
We present the temperature  multipole anisotropy in Fig.(\ref{fig:Tmod1},\ref{fig:Emod2}), the 
upper curve (gray) is the WMAP observation and the lower (black) one is our simulation. 
We find that the simulated temperature anisotropy is below the  WMAP data.
Our constrain on pseudoscalar-photon mixing is obtained from the 
E, B modes of CMBR. 

The above results correspond to spectral index $n_B =-2.37$, which is derived from the best 
fit of matter and CMBR power spectrum\cite{Yamazaki:2010nf}. Also the results are bound to 
the assumption that the background magnetic field has gone through a simple cosmological evolution. 
We have simulated the limits for $n_B =-2.20,-2.60$ and $-2.90$ also and observe a slight deviation in values. 
We present these results in table \ref{tb:nblist}.
\begin{table}[]
\centering
\begin{tabular}{|l|l|l|l|l| }
\hline
 $ m_\phi (eV)$     & \multicolumn{4}{c|}{$g_\phi B$($GeV^{-1} nG$)}      \\
                   & $n_B=-2.20$  & $n_B=-2.37$   & $n_B=-2.60$   &           $n_B=-2.90$ \\
\hline
$10^{-10}$ &  $1.55\times10^{-13}$  & $1.60\times10^{-13}$ & $1.70\times10^{-13}$  & $1.80\times10^{-13}$ \\
$10^{-15}$ &  $3.20\times10^{-15}$ & $3.40\times10^{-15}$  & $3.80\times10^{-15}$ & $4.80\times10^{-15}$ \\
\hline
\end{tabular}      
\caption{The effect of spectral index $n_B$ on $g_{\phi}B$ limits.}
\label{tb:nblist}
\end{table}

\begin{figure}[!t]
  \centering{
    \includegraphics[width=3.5in,angle=0]{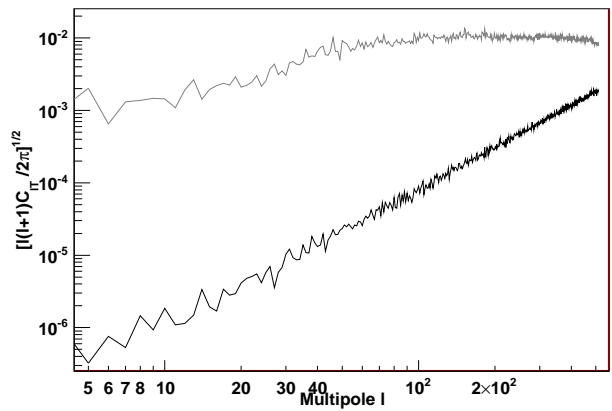}
    \caption{The simulated (black) and WMAP observed (gray) T mode  multipole, 
  $m_{\phi}=10^{-15}eV$, $g_{\phi}B =3.4\times10^{-15}GeV^{-1} nG$.}
\label{fig:Tmod2}}
\end{figure}
\section{Discussion}
\label{sc:dis}
We have done full $3D$ simulation of pseudoscalar-photon mixing for CMBR
at a very high resolution over the full sky. 
A comparison with WMAP observation results in a new and more stringent limit 
on the factor $g_{\phi}B$. It depends on the pseudoscalar mass and we 
simulate the limit on the factor $g_{\phi}B$  for two different masses of pseudoscalars.

Recently\cite{Horns:2012}, a bound on factor  $g_{\phi}B$  as 
$g_{\phi}B\le10^{-11}GeV^{-1}nG$  has been derived from 
ultraviolet photon polarization emerging from active galactic nuclei. Here the derived limit 
corresponds to ultra light ALPs($m_{\phi}\le10^{-15}eV$). 
In Ref.\cite{Mirizzi:2005,Mirizzi:2009}
the limits on $g_{\phi}B$ has been studied through CMBR spectral distortion, giving 
$g_{\phi}B\le10^{-13}\sim 10^{-11}GeV^{-1}nG$ for  ALPs masses between $10^{-15}eV$ and $10^{-4}eV$.
The pseudoscalar-photon mixing may also contribute to the  dimming of Type Ia 
supernovae\cite{Csaki:2002,Csaki:2002prl,Avgoustidis:2010}. The  phenomenon fixes  
$g_{\phi}B$ to $~10^{-11}GeV^{-1}nG$  for a axion of mass $10^{-16}eV$\cite{Csaki:2002prl}.

We may constrain $g_{\phi}$ form our bound on $g_{\phi}B$. However
the constrain on $g_{\phi}$ is subject to uncertainties in the background magnetic field.
Assuming the background magnetic field $B_0$ as $1nG$, our results bound $g_{\phi}\le1.6\times10^{-13}GeV^{-1}$ and
$g_{\phi}\le3.4\times10^{-15}GeV^{-1}$ for the ALPs of $10^{-10}eV$ and $10^{-15}eV$ respectively.

Our limits can be compared with  the direct experimental limits from SN1987A  , which is 
$g_{\phi}\le10^{-11}GeV$\cite{Brockway:1996} and $g_{\phi}\le3\times10^{-12}GeV$\cite{Grifols:1996} 
for very light ALPs ($\le10^{-9}eV$). We also recall the results from CAST\cite{Andriamonje:2007,Zioutas:2005},
$g_{\phi}\le8.8\times10^{-11}GeV$ for the ALPs of $0.02eV$, which of course is not for the ultralight 
ALPs and can not be directly compared with our results. 

We conclude that the CMBR multipole anisotropy imposes a stringent constraint on 
the pseudoscalar-photon coupling. We have obtained the lowest value
 of $g_{\phi}B$ as compared to available literatures.
\section*{Acknowledgements}
I acknowledge the use of the HEALPix \cite{Gorski:2005} software package. 
I am indebted to Prof. Pankaj Jain for many clarifying discussions and comments.
I also thank my friends and colleagues Gopal and Pranati for reading this 
draft and giving useful suggestions. I 
sincerely acknowledge CSIR, New Delhi for financial assistance in the 
form of Junior and  Senior Research Fellowship during the work.
\bibliographystyle{prsty}
\bibliography{CMBR}
\end{document}